\documentclass[prl,superscriptaddress,showpacs,showkeys,twocolumn,10pt]{revtex4-1}
\usepackage{amsfonts,bbold,amsmath,amssymb,graphicx,epstopdf,verbatim}
\usepackage[english]{babel}

\newcommand{\cc}{{\mathcal C}}

\begin{document}

\title{Measurement theory for closed quantum systems}
\author{Michiel Wouters}
\affiliation{TQC, Universiteit Antwerpen, Universiteitsplein 1, B-2610 Antwerpen, Belgium}
\date{\today}
\newcommand{\tr}{{\rm Tr}}
\begin{abstract}
We introduce the concept of a  ``classical observable'' as an operator with vanishingly small quantum fluctuations on a set of density matrices. It is shown how to construct them for a time evolved pure state. The study of classical observables provides a natural starting point to analyse the quantum measurement problem. In particular, it allows to identify Schr\"odinger cats and  the associated projection operators intrinsically, without the need to invoke an environment. We discuss how our new approach relates to the open system analysis of the quantum measurement problem.
\end{abstract}
\maketitle

`How does Hilbert space relate to our classical reality?' The consensus answer to this fundamental question is based on decoherence \cite{vonneumann,schlosshauer,joos,zurek_rmp}. Its starting point is a separation of the universe into system environment. The decoherence induced by the environment kills quantum effects and provides the sought-for transition from the quantum to the classical world.
While the separation of large systems into different parts clearly has its merits, the ambiguity in the division between system and environment remains awkward for a fundamental theory \cite{qsd,bell}. 
Currently, there is a thriving research activity on closed quantum systems \cite{polkovnikov}. In particular, much progress has been made in understanding their thermalisation \cite{rigol,reimann,short,polkov_diag}, a phenomenon where environments traditionally play a central role. We will show here how a theory of quantum measurement can be developed for closed systems.  

In order to define the relation between the quantum and classical worlds, we have to specify how the observable information relates to a quantum state. The crucial experimental fact is that all we know about the world are quantities with negligible quantum fluctuations. One may think here of an image on your computer screen, even though one does not have to go that far in the macroscopic world in practice. 
To make the connection between a quantum mechanical system and our knowledge about it, we therefore define the concept of a {\em classical observable} as an operator with vanishingly small quantum fluctuations. An experiment then corresponds to `reading' its expectation value. This procedure corresponds in all cases to experimental practice, where there is always a link between the quantum system and our `knowing it', that is described in terms of an expectation value and not in terms of a projection operator (think of the light emitted by the computer screen).  

We quantify the {\em classicality} of an operator with respect to a 
density matrix $\rho$ as
\begin{equation}
\cc(A)=\frac{[\tr ( A \rho )]^2}{\tr (A^2 \rho)}.
\label{eq:cc}
\end{equation}
It satisfies $\cc \leq 1$ because of the Cauchy-Schwarz inequality. For a classical observable, the upper limit is closely approached ($\cc \rightarrow 1$), where for an observable subject to large quantum fluctuations, $\cc$ is substantially smaller.
The unit operator clearly satisfies $\cc(\mathbb{1})=1$, expressing that the norm of the wave function does not fluctuate. For the density matrix of a pure state $\rho=|\psi \rangle \langle \psi |$, also the projection operator $P=|\psi \rangle \langle \psi |$ is classical. However, when the wave function is evolved in time, its classicality will in general quickly decrease.

\begin{figure}[hbtp]
\includegraphics[width=1\columnwidth]{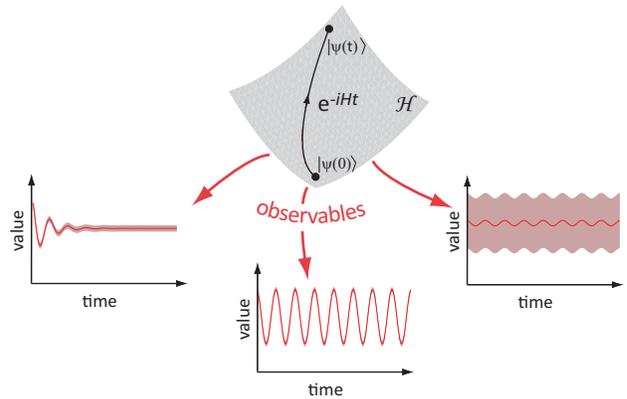}
\caption{Illustration of the dynamics of observables for a wave function that evolves in Hilbert space $\mathcal H$ under Hamiltonian evolution. Shaded areas indicate the magnitude of quantum fluctuations. Most observables have small signal to noise ratio and do not qualify as classical observables.}
\label{fig:0}
\end{figure}

In order to find operators that do remain classical under time evolution, we  maximize 
\begin{equation}
\overline{\cc}(A) = \frac{\int_0^T dt \{\tr [ A \rho(t) ]\}^2}{\int_0^T dt \tr [A^2 \rho(t)]}.
\label{eq:ccavg}
\end{equation}
The denominator can here be seen as a natural measure for the magnitude of the operator. For operators with zero time-averaged expectation value, there exists a simple relation between the classicality $\overline \cc$ and the signal to quantum noise ratio (see supplemental material for the precise definition and derivation):
\begin{equation}
{\rm SNR}(A) = \frac{1}{\sqrt{1-\overline \cc(A)}}.
\label{eq:snr}
\end{equation}
This relation tells us that classical observables, with $\overline \cc \to 1$, show temporal variations that are much larger than their fluctuations, while the time dependence of the other operators is drowned in noise (see Fig. \ref{fig:0}).

To complete the specification of the measurement problem, we have to choose an initial condition. If we only use the total Hamiltonian as an input for our analysis, we are restricted to formulate it in the energy eigenbasis $\vert n \rangle$. As an incoherent mixture of energy eigenstates precludes any dynamics, it is most natural to consider a pure state. We will assume that a finite number of $N$ energy eigenstates, in an energy window $[E,E+\Delta E]$ is populated. Because there is no dynamics in the populations of the eigenstates, we should not lose physics by making the simplifying assumption that the initial state has equal overlap with all the energy eigenstates within the energy window \cite{rigol,reimann}. We do not loose generality by taking the overlap  $\langle n \vert \psi(t=0) \rangle$ real for our specific initial condition, because the absolute phase of the energy eigenstates is arbitrary.

Let us start with the results for the harmonic case (for the derivation, see supplemental material), where the energy difference between the  states is constant. As expected, the two observables with the largest classicality $\overline \cc=1-1/N$ , are (approximately) the usual position ($X$) and momentum ($P$) operators, where
\begin{equation}
X = \sum_{n=1}^{N-1} \big(|n \rangle \langle n+1|+|n+1 \rangle \langle n|\big) .
\label{eq:X}
\end{equation}
For large $N$, also powers $X^n P^m$ are good classical observables for $n+m \ll N$, so that we can use ordinary calculus for sufficiently smooth functions of $X$ and $P$.
When the Hamiltonian is the direct product of two harmonic oscillator Hamiltonians $H=H_1 \bigotimes H_2$, both position and momentum operators are classical observables, as well as functions of them.

\begin{figure}[hbtp]
\includegraphics[width=1\columnwidth]{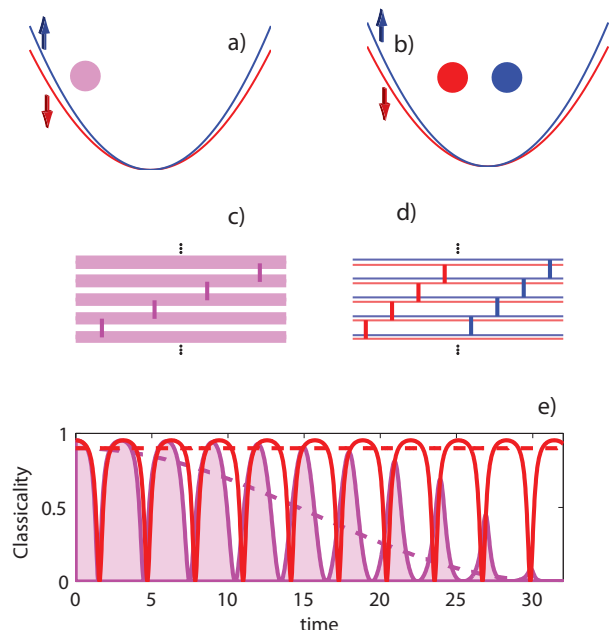}
\caption{a) Illustration of the qubit coupled to a harmonic oscillator potential. At early times (a), the positions in the two potentials (circles) are the same. It corresponds to the poor energy energy resolution, indicated by thick energy levels in panel c). All transitions have the same frequency within the Heisenberg resolution, allowing for the existence of a classical observable. At later times (panels b,d), when the two groups of transitions are dephased with respect to each other, Hilbert space `breaks up'. Only on the red and blue harmonic subspaces, classical observables exist. Panel (e) shows the instantaneous classicality of the $X$-operator, $\mathcal C(X)$ (full magenta), and $(\cc(X)+\cc(P))/2$ (dashed magenta). The red lines show the same for the density matrix restricted to the subspace $\mathcal H_1$. Because of the limited number of frequencies in our toy model, the classicalities show a revival at later times and are periodically repeated.}
\label{fig:1}
\end{figure}

The analysis becomes more interesting when we consider a direct sum of two incommensurate harmonic oscillator ladders. The total Hilbert space is then $\mathcal H= \mathcal H_1 \bigoplus \mathcal H_2$, each with $N/2$ levels, but with different energies: $\omega_n=n \omega_1$ for the first ladder and $\omega_n=n\omega_2$ for the second one (see Fig \ref{fig:1} c). It represents a qubit coupled to a harmonic oscillator, whose cavity-QED implementation has become a fruitful testing ground for quantum physics \cite{haroche,deleglise,vlastakis}. 
For short evolution times, $T\ll 2\pi/|\omega_1-\omega_2|$, the two harmonic oscillator ladders are indistinguishable within the Heisenberg limited energy resolution (see Fig. \ref{fig:1} d) and we find that $X=X_1\bigoplus X_2$ is still a classical observable, with $\overline \cc(X)=1-2/N$. 
The operators $X_1$ and $X_2$ are analogous to \eqref{eq:X}, but with the levels restricted to $\mathcal H_1$ and $\mathcal H_2$ respectively. 
For evolution times much longer than the Heisenberg uncertainty time 
$T\gg 2\pi/|\omega_1-\omega_2|$ on the other hand (panel d), a classical observable can no longer be found. The most classical ones are $X_1$ and $X_2$, with
$\overline \cc(X_1)=\overline \cc(X_2)=1/2-1/N$.

In general, the highest classicality is obtained for operators that consist of all resonant transitions $|n\rangle \langle m|$. High classicality up to time $T$ is obtained when all eigenstates form part of a harmonic oscillator ladder with common transition $\omega_c$, within the Heisenberg energy uncertainty $2 \pi/T$ (see Fig. \ref{fig:1} c). The corresponding classical observable is then fully collective: all states participate in it as in equation \eqref{eq:X}. The stringent requirements of harmonicity and collectivity provide an elementary explanation of the fact that classical phase space is so much smaller than Hilbert space.

The lack of a classical observable up to late times on the full Hilbert space $\mathcal H$ reflects the fact that the system turns into a Schr\"odinger cat. Note that we have identified the cat state without separating the universe in system and environment. Our analysis thus provides a solution to the `preferred-basis problem' \cite{schlosshauer,stapp,wallace} in Everett's relative-state approach \cite{everett}. 
Observing the expectation value of the $X$ at late times $T>2\pi/|\omega_1-\omega_2|$ will result in the collapse of the wave function. The density matrix is then projected to a single subspace $\mathcal H_1$ or $\mathcal H_2$, with a probability proportional to their respective dimensions. This projection is precisely what is needed to restore the classicality of the observable $X$, as illustrated in Fig. \ref{fig:1}. Where the classicality of the $X$-operator on the total Hilbert space (magenta lines) decays as a function of time, it remains constant on the projected Hilbert space (red lines).

\begin{figure}[hbtp]
\includegraphics[width=1\columnwidth]{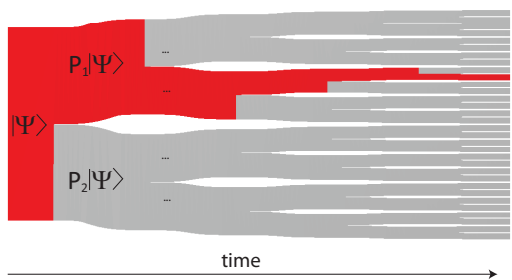}
\caption{As time goes on, observables are classical only for a wave function projected in progressively smaller subspaces of Hilbert space. The red region shows a consistent choice for consecutively selected subspaces. The grey regions are eliminated. }
\label{fig:2}
\end{figure}

For more complex systems, the breakup of Hilbert space will occur for different subspaces on different time scales, as schematically illustrated in Fig. \ref{fig:2}. At every branching point, a projection takes place. A possible scenario for surviving components is shown in red. Because the different branches are eigenstates of the full Hamiltonian, elimination is definitive. A `consistent history' \cite{griffiths,hartle,omnes} then appears naturally.

The connection with the open system decoherence approach, where the environment consists of harmonic oscillators \cite{schlosshauer,breupetr}, is straightforward.
In case the system itself has a classical observable, this remains a classical observable under linear coupling to the environment. The most familiar example is a single harmonic oscillator coupled harmonic to a harmonic bath \cite{breupetr,feynvern,caldlegg}.
On the other hand, if the system does not have a classical observable, the coupling with the environment is essential to have classicality at all. Classical observables can then be constructed for the projected system, combined with the environment. The required projection operators correspond to the usual pointer states \cite{pointer}.
Our Schr\"odinger cat example in Fig. \ref{fig:1} is the simplest illustration of this mechanism. For the qubit alone $\overline \cc(\sigma_{x,y})=1/2$ ($\sigma_{x,y}$ are the Pauli matrices) there is no classical observable. When it is coupled to a harmonic oscillator, the combined system does have classical observables after projection of the system on a pointer state. This mechanism can be generalized straightforwardly to multiple harmonic oscillators, as is the case in practice \cite{zurek_natphys,zurek_photon}, where the revival time (see caption of Fig. \ref{fig:1}) tends to infinity.

It is also worth discussing the connection to studies on thermalization in closed many-body quantum systems \cite{rigol,polkovnikov,reimann,short}. For generic many-body systems, it is expected that no harmonic oscillator ladders exist, i.e. that from $\omega_n-\omega_m=\omega_{n'}-\omega_{m'}$ it follows that either $n=m$ and $n'=m'$ or $n=n'$ and $m=m'$ \cite{reimann,peres,goldstein}. We then find that at late times, all classical dynamics with large SNR decay. Our investigation is thus complementary to the thermalization studies: where they are concerned with proving that most of the time for all observables the SNR is vanishingly small  \cite{reimann,short}, we focus on finding the observables that do show interesting dynamics in a given time window. 
Remember that the phase of the energy eigenstates was chosen such that the overlap 
$\langle n \vert \psi(t=0) \rangle$ was real. This can be done for any initial state, so that in every time window $[t,t+\tau_R]$, a similar set of classical observables can be constructed: if $A$ is a classical observable in the interval $[0,\tau_R]$, then the Heisenberg backward evolved operator $A'=e^{-iHt} A e^{iHt}$ is indeed trivially classical on the interval $[t,t+\tau_R]$. 
On a speculative note, the relation between a specific set of classical observables and a certain time window could mean that our classical experiences are in this sense related to the history of the universe. 

In conclusion, we have analysed quantum measurements of closed systems, based on the concept of `classical observable'. This is an operator with small quantum fluctuations and correspondingly large signal to noise ratio. We have shown that the spectrum of the Hamiltonian is sufficient to find all the classical observables. The condition for the existence of a classical observable up to time $T$ is that all states belong to a harmonic oscillator ladder within an energy resolution $2\pi/T$. Projection restores classicality when it ceases to exist for the full wave function. The physical results of our approach coincide with the ones from standard quantum measurement theory, as developed for open systems, but we believe that this different perspective is conceptually clarifying and hope that it may offer guidance to address open problems.

I gratefully acknowledge stimulating discussions with Dries Sels and Jacques Tempere. This work was financially supported by the FWO Odysseus program.



\newpage
\def\theequation{S.\arabic{equation}}
\setcounter{equation}{0}

{\section{Supplemental information}

We maximize the classicality $\overline \cc$, Eq. \eqref{eq:ccavg} in the main text, by expanding the operator $A$ in a basis $B_i$ for the linear Hermitian operators acting on the Hilbert space: $A = \sum_i a_i B_i$. 
For the eigenvectors of the generalized eigenvalue problem
\begin{equation}
\sum_j R_{ij} a^{(n)}_j = \overline \cc_n \sum_j M_{ij} a^{(n)}_j,
\label{eq:geneig}
\end{equation}
the classicality corresponds to the generalized eigenvalue $\overline \cc_n$ . 
In Eq. \eqref{eq:geneig} The matrices $R_{ij}$ and $M_{ij}$ are symmetric and defined as
\begin{align}
R_{ij} &= \frac{1}{T}\int_0^T dt \; \tr[B_i \rho(t)]\tr[B_j \rho(t)] 
\label{eq:defR}
\\
M_{ij} &= \frac{1}{2} \tr[ (B_i B_j +B_j B_i) \; \overline{\rho} ].
\label{eq:defM}
\end{align}
Two operator eigenvectors $A_n=\sum_i a^{(n)}_i B_i$ of Eq. \eqref{eq:geneig} that belong to different eigenvalues are orthogonal in the following sense:
\begin{equation}
 \frac{1}{2} \tr[ (A_m A_n +A_n A_m) \; \overline{\rho} ] = 0.
\end{equation}
Because the unit operator is a generalized eigenvector, the other operators have all zero time averaged expectation value: $\tr(A_m \overline \rho)=0$ (if there are other operators with $\overline \cc=1$, they can be chosen to be so). 

For an operator with zero expectation value, the square of the expectation value is a good measure of its signal. This motivates us to define the signal to noise ratio (SNR) as
\begin{equation}
{\rm SNR}(A) = \sqrt{\frac{\int_0^T dt \{\tr[A \rho(t)]\}^2}{\int_0^T dt \tr[A^2 \rho(t)]- \{ \tr[A \rho(t)]\}^2}}.
\end{equation}
With the definition of the classicality (Eq. \eqref{eq:cc} in the main text), we immediately obtain Eq. \eqref{eq:snr}.

As a basis for the Hermitian operators on the Hilbert space, we choose
\begin{align}
\label{eq:x}
x_{m,n} &= \vert m \rangle \langle n \vert +\vert n \rangle \langle m \vert 
\\ \label{eq:p} 
p_{m,n} &= -i(\vert m \rangle \langle n \vert -\vert n \rangle \langle m \vert),
\end{align}
with the $m\neq n$ restricted to the energy levels that are populated according to the initial condition.

Because quantum mechanics is defined on projective Hilbert space (the absolute phase of a state does not matter) \cite{breupetr}, the phase freedom can be used to make the overlap  $\langle k \vert \psi(t=0) \rangle$ real. We then have for the density matrix
\begin{equation}
\rho (t) = \frac{1}{N} \sum_{k,l=1}^N e^{-i t(\omega_k -\omega_l)} \vert k \rangle \langle l \vert .
\label{eq:rho}
\end{equation}

We then have the following expectation values
\begin{align}
\tr[x_{m,n} \rho(t)] = 2 \cos[(\omega_m-\omega_n)t] \label{eq:trx} \\
\tr[p_{m,n} \rho(t)] = 2 \sin[(\omega_m-\omega_n)t] 
\end{align}
For sufficiently long times $T \gg 2 \pi \,  {\rm max}(|\omega_m-\omega_n|^{-1})$ the matrix elements $R_{ij}$, where $i$ is of the $x$-type and $j$ is of the $p$ type operator vanish. We can therefore restrict our search to operators of the $x$-type, and will automatically find a corresponding $p$-type operator.
For long times, we find that the matrix $M$ in \eqref{eq:defM} is proportional to the unit matrix: $M=(2/N) \mathbb{1}$. Degeneracies are discussed in the end.

The elements of the matrix $R$, restricted to the $x$-operator space can be then written more explicitly as $R_{mn,m'n'}$. Equations \eqref{eq:defR} and \eqref{eq:trx} then show that off-diagonal elements of $R$ vanish when the transition frequencies differ much more than the Heisenberg energy uncertainty, i.e. if $(|\omega_n-\omega_m|-|\omega_{n'}-\omega_{m'}|) \gg 2\pi/T$. The matrix $R$ is thus approximately block-diagonal, with each block corresponding to some transition frequency. Within a block of size $N_b$, all the matrix elements are equal to $2/N^2$.

The eigenvalue problem \eqref{eq:geneig} has a single nonzero eigenvalue per block in $R$, with corresponding eigenvector in the block $V=(1,1, \ldots,1)^T$ (all transitions act in phase). Because $RV=2 N_b/N^2V$ and $MV=(2/N)V$, we obtain $\overline \cc=N_b/N$. For a harmonic ladder, we have $N_b=N-1$, so that we find 
\begin{equation}
\overline \cc (X)=\overline \cc (P)=1-\frac{1}{N},
\label{eq:cc_ho}
\end{equation}
where the operators $X$ and $P$ are given by 
\begin{align}
X &= \sum_{n=1}^{N-1} \big(|n \rangle \langle n+1|+|n+1 \rangle \langle n|\big) \\
P &= -i\sum_{n=1}^{N-1} \big(|n \rangle \langle n+1|-|n+1 \rangle \langle n|\big) 
\label{eq:defX}
\end{align}
As expected, we have approximately recovered the harmonic oscillator position and momentum operators, that can be made arbitrarily classical by increasing the number of levels. For a two-level system $N=2$, we simply recover the Pauli matrices and find that $\overline \cc(\sigma_x)=\overline \cc(\sigma_y)=1/2$.
The only nonzero elements of the commutator between position and momentum operators are  $([X,P])_{1,1}=-([X,P])_{N,N}=2i$. 
In the limit of a large number of levels $N\gg 1$, the commutator is thus negligible with respect to $X$ and $P$.
Apart from the operators $X$ and $P$, their powers have classicality $\overline \cc(X^n)=\cc(P^n)=1-1/(N-n)$. As long as $n \ll N$, we find that $X^n$ and $P^n$ are classical operators. Ordinary calculus can thus be used for sufficiently smooth functions of $X$ and $P$.

Note that the initial state $| \psi_0 \rangle = 1/\sqrt{N} \sum_k |k \rangle $ is special with respect to the observable $X$, because then all $N-1$ terms in $X$ add up constructively in the expectation value. $\langle X \rangle$ then takes its maximal value
$\langle \psi_0 \vert X \vert \psi_0 
\rangle = 2-2/N$. When this maximal value is observed, this puts a severe restriction on the phases of the wave function. This means that the maximal value of $X$ should be a state of low entropy. A more elaborate analysis of entropy evolution along these lines thus looks promising.

The fact that we do not precisely recover the usual harmonic oscillator position and momentum operators (a factor $\sqrt{n+1}$ is missing in Eq. \eqref{eq:defX}) can be attributed to the fact that the energy distribution is not Poissonian, but uniform in the interval $[E,E+\Delta E]$ (see discussion above Eq. \eqref{eq:rho} in the main text). This difference also causes the breakdown of the usual commutation relation $[X,P]=i \hbar $. Note that in the limit of a small energy window $\Delta E \lll E$ at high energy $E \gg \omega$, the variation of $\sqrt{n+1}$ becomes negligible and $X$ approaches the usual harmonic oscillator positon operator.

When the states have a degeneracy $g$, the matrix $M$ is no longer diagonal, but it consists of the blocks of transitions that each have degeneracy $g$ (within Heisenberg resolution). We then find $M V= (2/N) g V$, so that the classicality equals $\overline \cc = N_b/(N g) $. Because $N_b=g^2 N_l$ and $N=g N_l$ ($N_l$ is the number of distinct levels), we then have that for harmonically spaced, degenerate levels that $\overline \cc=1-1/N_l$ .

\end{document}